\documentclass{article}
\usepackage{amssymb}
\usepackage{amsmath}
\usepackage{amsfonts}
\usepackage{authblk}
\usepackage{enumerate}
\def\Ric{\mathop{\rm Ric}\nolimits}
\begin{document}
\title{Integrability of geodesics and action-angle variables in 
Sasaki-Einstein space $T^{1,1}$}

\author{Mihai Visinescu\thanks{mvisin@theory.nipne.ro}}

\affil{Department of Theoretical Physics,

National Institute of Physics and Nuclear Engineering,

Magurele, P.O.Box M.G.-6, Romania}

\date{} 
\maketitle

\begin{abstract}

We briefly describe the construction  of St\"{a}\-kel-Killing and Killing-Yano
tensors on toric Sasaki-Einstein manifolds without working out intricate 
generalized Killing equations. The integrals of geodesic motions are expressed 
in terms of Killing vectors and Kill\-ing-Yano tensors of the 
homogeneous Sasaki-Einstein space $T^{1,1}$. We discuss the integrability of 
geodesics and construct explicitly the action-angle variables.
Two pairs of frequencies of the geodesic motions are resonant giving way to
chaotic behavior when the system is perturbed.

~


~


~

\end{abstract}

\section{Introduction}

The renewed interest in Sasaki geometries has arisen in connection with some 
recent developments in mathematics and theoretical physics \cite{JS}. Sasakian 
geometry represents the natural odd-dimensional counterpart of the K\"{a}hler 
geometry. Sasaki structure in $(2n-1)$ dimensions is sandwiched between the 
K\"{a}hler metric cone in $n$ complex dimensions and the transverse 
K\"{a}hler structure of complex dimension $(n-1)$. In particular the K\"{a}hler 
cone is Ricci flat, i.e. Calabi-Yau manifold, if and only if the corresponding 
Sasaki manifold is Einstein.

The interest in physics for Sasaki-Einstein (SE) geometry is connected with 
its relevance in AdS/CFT duality. New nontrivial infinite family of toric SE manifolds 
were constructed \cite{GMSW} and  many new insights were obtained for AdS/ CFT 
correspondence. These SE spa\-ces are denoted by $Y^{p,q}$ and characterized by the
two coprime positive integers $p$ and $q$ with $q<p$. A vastly greater number of
SE spaces was constructed in \cite{CLPP} and denoted by $L^{p,q,r}$ where $p,q$ and $r$
are coprime positive integers with $0 < p \leq q\,, 0 < r <p+q$ and with $p$ and $q$
each coprime to $r$ and to $s=p +q-r$. These metrics have $U(1)\times U(1) \times U(1)$
isometry. In the special case $p+q=2 r$ the isometry of these metrics is enlarged to
$SU(2)\times U(1) \times U(1)$ which is the isometry of the spaces $Y^{p,q} = 
L^{p-q, p+q,p}$. Another special limit is $p=q=r=1$ and the metric becomes the 
homogeneous $T^{1,1}$ space with $SU(2)\times SU(2) \times U(1)$ isometry.

The symmetries of SE spaces play an important role in connection with the study of 
integrability properties of geodesic motions and separation of variables of the 
Hamilton-Jacobi or quantum Klein-Gordon, Dirac equations.

Higher order symmetries associated with St\"{a}ckel-Killing (SK) and Killing-Yano (KY)
tensors generate conserved quantities polynomial in momenta describing the so called 
\emph{dynamical} or \emph{hidden symmetries}. In general it is a difficult task to solve
straightly the generalized Killing equations satisfied by KY and SK tensors. Using the 
geometrical structure of a SE manifold, its connection with the complex structure of the 
Calabi-Yau metric cone it is possible to produce explicitly the complete set of Killing 
tensors and consequently the integrals of the geodesic motions.

One of the interesting aspect of AdS/CFT correspondence is integrability which allows 
obtaining many new classical solutions of the theory (see the review \cite{NB}). By 
focusing on the integrability of the geodesics on SE spaces we gain a better 
understanding of the geometries produced by $D$-branes  on non-flat bases.

The standard way to decide if a system is integrable is to find integrals of motion.
A dynamical system is called integrable if the number of functionally independent 
integrals of motion is equal to the number of degrees of freedom.

In the context of AdS/CFT duality type $IIB$ strings on $AdS_5 \times S^5$ with 
Ramond-Ramond fluxes is related to ${\cal N} = 4 \; SU(N)$ gauge theory. An interesting 
generalization of this duality between gauge theory and strings is to consider backgrounds 
of the form $AdS_5 \times X_5$ where $X_5$ is in a general class of five-dimensional
Einstein spaces admitting $U(1)$ fibration. While the type $IIB$ string theory on 
$AdS_5 \times S^5$ is classically integrable \cite{BPR} there are many non-integrable 
AdS/CFT dualities in which the string world-sheet theory exhibits a chaotic behavior.
This is the case when the internal space is a SE manifold like $T^{1,1}$ or $Y^{p,q}$
\cite{BZ2}. 

The purpose of this paper is to analyze the integrability of geodesics of the homogeneous 
regular SE $5$-dimensional space $T^{1,1}$. In the light of the AdS/CFT correspondence,
$AdS \times T^{1,1}$ represents the first example of a supersymmetric holographic 
theory based on a compact manifold which is not locally $S^5$ \cite{KW}.

Using the multitude of Killing vector (Kv) fields and SK tensors it is possible to construct 
the conserved quantities for geodesic motions on $T^{1,1}$. However, the number of 
functionally independent constants of motion is only $5$, implying the complete integrability 
of geodesics, but not superintegrability.

In order to understand the peculiarities of the geo\-de\-sic motions in $T^{1,1}$ space 
we shall perform the \emph{action-angle} formulation of the phase space. The description of 
the integrability of geodesics in $T^{1,1}$ in these variables gives us a 
comprehensive geometric description of the dynamics.

The existence of the action-angle variables is very important both for the theory of 
near-integrable systems (KAM theory) and for the quantization of integrable systems 
(Bohr-Sommerfeld rule). The action-angle variables define an $n$-dimensional surface 
which is a topological torus (Kolmogorov-Arnold-Moser (KAM) tori) \cite{VIA}.
The KAM theorem states that when an integrable Hamiltonian is perturbed by a small 
nonintegrable piece most tori survive but suffer small deformations. On the other hand
the resonant tori which have rational ratios of frequencies get destroyed and motion on
them becomes chaotic. The use of action-angle variables provides a powerful technique 
to quickly obtain the frequencies of the periodic motions without finding a complete 
solution to the motion of the system.

The paper is organized as follows. In the next two Sections we recall some definitions and 
known results concerning the Killing tensors and SE spaces. For the completeness and clarity 
of the exposition, in Section 4 we present the constants 
of motion on $T^{1,1}$ space proving the integrability of geodesics. In Section 5 we construct 
explicitly the action-angle variables and the frequencies of the motions. Finally in Section 6 
we present some concluding remarks and discuss the issues regarding the presence of resonant 
frequencies. For the convenience of the reader, some details concerning the evaluation of 
some integrals from Section 5 are deferred to the Appendix.
 
\section{St\"{a}ckel-Killing and Killing-Yano tensors}

SK and KY tensors stand as a natural extension of the Kv fields which are linked to the 
continuous isometries that leave the metric invariant.

On a Riemannian manifold $(M,g)$ with local coordinates $x^\mu$ and metric $g_{\mu\nu}$
the geodesics can be obtained as the trajectories of test-particles with proper-time
Hamiltonian
\begin{equation}\label{Ham}
H = \frac12 g^{\mu\nu} p_\mu p_\nu\,.
\end{equation}

In the presence of a Kv the system of a
free particle with Hamiltonian \eqref{Ham} admits a conserved quantity
linear in the momenta.

A SK tensor of rank $r$ is a symmetric tensor defined on the manifold $M$
\begin{equation}
K = \frac{1}{r!} K^{(\mu_1 \cdots \mu_r)} \frac{\partial}{\partial x^{\mu_1}}
\otimes \cdots \otimes \frac{\partial}{\partial x^{\mu_r}}
\end{equation}
such that 
\begin{equation}\label{SK}
\nabla_{(\mu}  K_{\mu_1 \cdots \mu_r)} =0 
\end{equation}
where $\nabla$ is the Levi-Civita connection.
If $M$ admits a SK tensor, the system of a free particle possesses a conserved quantity 
of higher order in the momenta
\begin{equation}\label{CSK}
C_{SK} =  K^{(\mu_1 \cdots \mu_r)} p_{\mu_1}\cdots p_{\mu_r} \,.
\end{equation}

A different generalization of the Kv's is represented by the KY tensors. 
A KY tensor is a differential $p$-form defined on $M$ 
\begin{equation}
\Psi = \frac{1}{r!} \Psi_{[\mu_1 \cdots \mu_r]} dx^{\mu_1} \cdots dx^{\mu_r}
\end{equation}
satisfying the equation
\begin{equation}\label{KY}
\nabla_{(\mu} \Psi_{\mu_1)\cdots \mu_r} =0\,. 
\end{equation}

Giving two KY tensors  $\Psi^{\mu_1, \dots, \mu_r}$ and
$\Sigma^{\mu_1, \dots, \mu_r}$ the partial contracted product generates a 
SK tensor of rank $2$:
\begin{equation}\label{KYY}
K^{(\Psi,\Sigma)}_{\mu\nu} = \Psi_{\mu \lambda_2 \dots \lambda_r}
\Sigma_{\nu}^{\phantom{j} \lambda_2 \dots \lambda_r}+ 
\Sigma_{\mu \lambda_2 \dots \lambda_r}
\Psi_{\nu}^{\phantom{j} \lambda_2 \dots \lambda_r }\,.
\end{equation}
This property offers a method to generate higher order integrals of motion \eqref{CSK}
by identifying the complete set of KY tensors.

It is worth mentioning that in general is is a hard task to find the complete set of SK 
or KY tensors trying to solve directly eqs. \eqref{SK}, \eqref{KY}. In some cases it is
possible to produce the complete set of KY tensors taking advantage of geometrical 
properties of the space. That is the case of toric SE spaces for which
the explicit construction of KY tensors is possible \cite{US}. 

\section{Sasaki-Einstein spaces}

Recall that a $(2n-1)$-dimensional manifold $M$ is a \emph{contact manifold} if there exists
a $1$-form $\eta$, called a contact $1$-form, on $M$ such that 
\begin{equation}
\eta \wedge (d \eta)^{n-1} \neq 0
\end{equation}
everywhere on $M$ \cite{B-G-2008}. For every choice  of contact $1$-form $\eta$ there
exists a unique vector field $K_{\eta}$, called Reeb vector field, that satisfies
\begin{equation}\label{Rv}
\eta (K_{\eta}) = 1 \quad \text{and} \quad K_{\eta} \lrcorner d\eta = 0\,.
\end{equation}
The Reeb vector field $K_{\eta}$ is a Kv field on $M$, has unit length and
its integral curve is a geodesic.

A contact Riemannian manifold is Sasakian if its metric cone
\begin{equation}
(C(M), \bar{g}) = (\mathbb{R}_+ \times M, dr^2+r^2g)
\end{equation}
is K\"{a}hler. Here $r \in (0,\infty)$ may be considered as a coordinate on the
positive real line $\mathbb{R}_+$.

As part of the connection between Sasaki and K\"{a}h\-ler geometries it is worth 
noting that in the case of a SE manifold, $\Ric_g = 2(n-1) g$, the metric
cone is Ricci flat $\Ric_{\bar{g}} = 0$, i.e. a Calabi-Yau manifold.

The KY tensors on a SE manifold are described by the 
Killing forms
\begin{equation}\label{Psik}
\Psi_k = \eta \wedge (d \eta)^k \quad, \quad k = 0,1, \cdots , n-1\,.
\end{equation}

These KY tensors do not exhaust the complete set of KY tensors on SE spaces. 
The Calabi-Yau metric cone has holonomy $SU(n)$ and admits two additional
parallel forms given by the real and imaginary parts of the complex volume form 
\cite{US}. In order to complete the set of KY tensors on a SE manifold it is
necessary to use the relation between the KY tensors on SE space and its metric cone.
More precisely, for any $p$-form $\Psi$ on the space $M$ we can define an associated
$(p+1)$-form $\Psi^C$ on the cone $C(M)$ \cite{US}.

\section{Integrability of geodesics in Sasaki-Einstein space $T^{1,1}$}

Any complete homogeneous SE $5$-dimensional manifold is a $U(1)$-bundle over the 
complex projective plane $\mathbb{C} P^2$ or $\mathbb{C} P^1\times \mathbb{C} P^1$
\cite{JS}. The well known realizations are the round metric on $S^5$ and the
homogeneous metric $T^{1,1}$ on $S^2\times S^3$.

The metric on $T^{1,1}$ is \cite{CO,MS}
\begin{equation}\label{mT11}
\begin{split}
ds^2(T^{1,1}) =&  
\frac16 (d \theta^2_1 + \sin^2 \theta_1 d \phi^2_1 +
d \theta^2_2 + \sin^2 \theta_2 d \phi^2_2) \\
&+ \frac19 (d \psi + \cos \theta_1 d \phi_1 + \cos \theta_2 d \phi_2)^2 \,.
\end{split}
\end{equation}
Here $\theta_i\, , \phi_i, \quad i=1,2$ are the usual coordinates on two round $S^2$ 
spheres and $\psi \in [0, 4 \pi)$ parametrizes the $U(1)$ fiber over $S^2\times S^2$.

The globally defined contact $1$-form $\eta$ is:
\begin{equation}\label{etaT}
\eta =\frac13 (d \psi +\cos \theta _1 d \phi _1+\cos \theta _2 d \phi _2)
\end{equation}
and the corresponding Reeb vector is 
\begin{equation}\label{TRv}
K_{\eta} = 3 \frac{\partial}{\partial \psi}\,.
\end{equation}
In what follows we define $2\nu = \psi$ so that $\nu$ has canonical period $2\pi$.

On the manifold $T^{1,1}$ with the metric \eqref{mT11} the geo\-desics are described 
by the Hamiltonian \eqref{Ham} where the canonical momenta conjugate to the coordinates
$x^\mu$ are $p_\mu = g_{\mu\nu}\dot{x}^\nu$ with overdot denoting proper time derivative.
In particular the conjugate momenta to the coordinates 
$(\theta_1,\theta_2,\phi_1,\phi_2,\nu)$ are:
\begin{equation}
\begin{split}
p_{\theta_1} =& \frac16 \dot{\theta}_1\\
p_{\theta_2} =& \frac16 \dot{\theta}_2\\
p_{\phi_1} =& \frac16 \sin^2\theta_1 \,\dot{\phi}_1 + 
\frac19 \cos^2\theta_1 \,\dot{\phi}_1
+ \frac29 \cos\theta_1 \,\dot{\nu} \\
&+\frac19 \cos\theta_1 \cos\theta_2\,\dot{\phi}_2\\
p_{\phi_2} =& \frac16 \sin^2\theta_2 \,\dot{\phi}_2 + 
\frac19 \cos^2\theta_2 \,\dot{\phi}_2
+ \frac29 \cos\theta_2 \,\dot{\nu} \\
&+ \frac19 \cos\theta_1 \cos\theta_2 \,\dot{\phi}_1\\
p_{\nu} =& \frac29 \,\dot{\nu} + \frac19\cos\theta_1 \,\dot{\phi}_1 + 
\frac19\cos\theta_2 \,\dot{\phi}_2 
\end{split}
\end{equation}
and the conserved Hamiltonian \eqref{Ham} takes the form:
\begin{equation}\label{HT11}
\begin{split}
 H=& 3 \left[ p^2_{\theta_1} +  p^2_{\theta_2} +
\frac{1}{4\sin^2\theta_1}(2 p_{\phi_1} - \cos\theta_1 p_{\nu})^2 \right.\\
&\left. +\frac{1}{4\sin^2\theta_2}( 2 p_{\phi_2} - \cos\theta_2 p_{\nu})^2 
\right] 
 + \frac98 p^2_{\nu} \,.
\end{split}
\end{equation}

Taking into account the isometries of $T^{1,1}$, momenta $p_{\phi_1},p_{\phi_2}$ 
and $p_{\nu}$ are conserved. Since $T^{1,1}$ has $SU(2)\times SU(2)\times U(1)$
symmetry, two total $SU(2)$ angular momenta are also conserved:
\begin{equation}
\begin{split}
\vec{J}_1^{~2} =& p_{\theta_1}^2 + \frac{1}{4\sin^2\theta_1}( 2 p_{\phi_1}
- \cos\theta_1 p_{\nu})^2  + \frac14 p^2_{\nu} \\
\vec{J}_2^{~2} =& p_{\theta_2}^2 + \frac{1}{4\sin^2\theta_2}(2 p_{\phi_2}
- \cos\theta_2 p_{\nu})^2  + \frac14 p^2_{\nu} \,.
\end{split}
\end{equation}

Other constants of motion are constructed according to \eqref{KYY} from the KY 
tensors of $T^{1,1}$. Firstly there are two KY tensors \eqref{Psik} for $k=1,2$ 
constructed from the contact form $\eta$ \eqref{etaT}. Finally there are two
additional KY tensor related to the complex volume form of the metric cone
$C(T^{1,1})$. All these constants of motion are explicitly given in 
\cite{BV,S-V-V1}.

In spite of the existence of a multitude of constants of motion, the number of
functionally independent is only $5$ \cite{BV,S-V-V2}, exactly the dimension of 
the SE space $T^{1,1}$. In the next Section we shall use the complete integrability
of geodesics to solve the Hamilton-Jacobi equation by separation of variables and 
construct the action-angle variables.

\section{Action-angle variables}

We start with the Hamilton-Jacobi equation
\begin{equation}
H(q_1,\cdots,q_n; \frac{\partial S}{\partial q_1},\cdots,
\frac{\partial S}{\partial q_n}; t) + \frac{\partial S}{\partial t} =0 
\end{equation}
with the \emph{Hamilton's principal function}
\begin{equation}
S=S(q_1,\cdots,q_n;\alpha_1,\cdots,\alpha_n;t)
\end{equation}
depending on $n$ constants of integration. Taking into account that the 
Hamiltonian \eqref{HT11} has no explicit time dependence, Hamilton's 
principal function can be written in the form \cite{GPS}
\begin{equation}
S(q,\alpha,t) = W(q,\alpha) - E t
\end{equation}
where one of the constants of integration is equal to the constant value 
$E$ of the Hamiltonian and $W(q,\alpha)$ is the \emph{Hamilton's 
characteristic function} 
\begin{equation}\label{sv}
W = \sum \int p_i \dot{q}_i dt = \int p_i d q_i\,.
\end{equation}

If the motion stays finite, each $q_i$ will go through repeated cycles and 
the canonical \emph{action variables} are defined as integrals over complete 
period of the orbit in the $(q_i,p_i)$ plane
\begin{equation}
J_i = \oint p_i d q_i = \oint \frac{\partial W_i(q_i;\alpha)}{\partial q_i}
dq_i \qquad \text{(no summation)}\,.
\end{equation}

$J_i$'s form $n$ independent functions of the constants $\alpha_i$ and can 
be taken as a set of new constant momenta.

Conjugate \emph{angle variables} $w_i$ are defined by the equ\-ations:
\begin{equation}\label{av}
w_i = \frac{\partial W}{\partial J_i} = 
\sum_{j=1}^n \frac{\partial W_j(q_j;J_1,\cdots,J_n)}{\partial J_i}
\end{equation}
having a linear evolution in time
\begin{equation}
w_i = \omega_i t + \beta_i 
\end{equation}
with $\beta_i$ other constants of integration and $\omega_i$ are frequencies
associated with the periodic motion of $q_i$.

In the case of geodesic motions in $T^{1,1}$ space the variables in the 
Hamilton-Jacobi equation are separable and we seek a solution with the characteristic 
function
\begin{equation}\label{W}
\begin{split}
W(\theta_1,\theta_2,\phi_1,\phi_2,\nu) =& W_{\theta_1}(\theta_1) + 
W_{\theta_2}(\theta_2) + W_{\phi_1}(\phi_1)\\
&+  W_{\phi_2}(\phi_2)
+  W_{\nu}(\nu)\,.
\end{split}
\end{equation}

Since $\phi_1,\phi_2,\nu$ are cyclic variables we have
\begin{equation}\label{cyc}
\begin{split}
& W_{\phi_1} = p_{\phi_1} \phi_1 = \alpha_{\phi_1} \phi_1\\
& W_{\phi_2} = p_{\phi_2} \phi_2 = \alpha_{\phi_2} \phi_2\\
& W_{\nu} = p_{\nu} \nu = \alpha_{\nu} \nu 
\end{split}
\end{equation}
where $\alpha_{\phi_1}, \alpha_{\phi_2}, \alpha_{\nu}$ are constants of 
integration.

The action variables corresponding to the cyclic coordinates can be straightly
obtained from \eqref{cyc}
\begin{equation}
\begin{split}
&J_{\phi_1} = 2 \pi \alpha_{\phi_1} \\
&J_{\phi_2} = 2 \pi \alpha_{\phi_2} \\
&J_{\nu} = 2 \pi \alpha_{\nu} \,.
\end{split}
\end{equation}

Using \eqref{cyc} the Hamilton-Jacobi equation reduces to
\begin{equation}\label{avtheta}
\begin{split}
E=
&\quad 3\Biggl[\biggl(\frac{\partial W_{\theta_1}}{\partial \theta_1}\biggr)^2 
+ \frac{1}{4\sin^2\theta_1} \bigl(2\alpha_{\phi_1} 
-\cos\theta_1 \alpha_\nu\bigr)^2\Biggr]\\
&+ 3\Biggl[\biggl(\frac{\partial W_{\theta_2}}{\partial \theta_2}\biggr)^2 
+\frac{1}{4\sin^2\theta_2} \bigl(2\alpha_{\phi_2} 
- \cos\theta_2 \alpha_\nu\bigr)^2\Biggr]\\
&+ \frac98 \alpha^2_\nu \,.
\end{split}
\end{equation}

We observe that the dependences on $\theta_1$ and respectively $\theta_2$ has
been separated into the expressions within the square brackets. The quantities 
in the square brackets must be constants which will be denoted by 
$\alpha^2_{\theta_1}$ and $\alpha^2_{\theta_2}$.

To find the action variables $J_{\theta_i}\,, (i=1,2)$ we infer from 
\eqref{avtheta}:
\begin{equation}\label{Wti}
\frac{\partial W_{\theta_i}}{\partial \theta_i} =
\sqrt{\alpha^2_{\theta_i} - \frac{(2\alpha_{\phi_i} - \cos\theta_i 
\alpha_\nu)^2}
{4\sin^2\theta_i}} \quad,\quad i=1,2 
\end{equation}
and the corresponding action variables are
\begin{equation}
J_{\theta_i} = \oint d\theta_i \sqrt{\alpha^2_{\theta_i} - 
\frac{(2\alpha_{\phi_i} - 
\cos\theta_i \alpha_\nu)^2}
{4\sin^2\theta_i}} \quad,\quad i=1,2 \,.
\end{equation}

The limits of integrations are defined by the roots $\theta_{i-}$ and
$\theta_{i+}$ of the expressions in the square root signs and a complete
cycle of $\theta_i$ involves going from $\theta_{i-}$ to $\theta_{i+}$
and back to $\theta_{i-}$.

This integral can be evaluated by elementary means or using the complex integration 
method of residues. In the later case, putting $\cos \theta_i = t_i$, for the 
evaluation of the action variables $J_{\theta_i}$, we extend $t_i$ to a complex variable 
$z$ and interpret the integral as a closed contour integral in the complex $z$-plane.
The turning points of the $t_i$-motion are
\begin{equation}
t_{i\pm} = 2\frac{\alpha_{\phi_i}\alpha_\nu \pm 
\alpha_{\theta_i}\sqrt{4\alpha^2_{\theta_i} + 
\alpha^2_\nu -4\alpha^2_{\phi_i}}}{4\alpha^2_{\theta_i} + \alpha^2_\nu}
\end{equation}
which are real for
\begin{equation}\label{constr}
4\alpha^2_{\theta_i} + \alpha^2_\nu 
-4\alpha^2_{\phi_i}\geq 0 
\end{equation}
and situated in the interval $(-1,+1)$. We cut the complex $z$-plane from $t_{i-}$ to 
$t_{i+}$ and the closed contour integral of the integrand is a loop enclosing the cut 
in a clockwise sense. The contour can be deformed to a large circular contour plus two 
contour integrals about the poles at $z= \pm 1$. After simple evaluation of the residues 
and the contribution of the large contour integral we finally get:
\begin{equation}
J_{\theta_i} = 2\pi\Biggl[\frac12\sqrt{4\alpha^2_{\theta_i} + \alpha^2_\nu} - 
\alpha_{\phi_i} \Biggr] \quad,\quad i=1,2 \,.
\end{equation}

We notice that the constants $J_{\theta_i}, J_{\phi_i}, J_\nu$ and $E$
are connected by the relation:
\begin{equation}\label{HTJ}
H = E = \frac{3}{4\pi^2}\biggl[\bigl(J_{\theta_1} + J_{\phi_1}\bigr)^2 +
\bigl(J_{\theta_2} + J_{\phi_2}\bigr)^2 -\frac18 J^2_\nu \biggr]
\end{equation}
making manifest that the Hamiltonian depends only on the action variables.
The number of independent constants of motion is {\it five} implying the complete 
integrability of geodesics on $T^{1,1}$.

A particular advantage of the change to action-angle variables is that one can 
identify the fundamental frequencies of the system. In this regard let us observe 
that the Hamiltonian \eqref{HTJ} depends on $J_{\theta_i}$ and $J_{\phi_i}$ 
in the combination $J_{\theta_i} + J_{\phi_i}$ meaning that the frequencies of 
motion in $\theta_i$ and $\phi_i$ are identical:
\begin{equation}\label{freq}
\omega_{\theta_i} = \omega_{\phi_i}= \frac{\partial H}{\partial J_{\theta_i}}=
\frac{\partial H}{\partial J_{\phi_i}}=\frac3{2\pi^2}\bigl(J_{\theta_i}+
J_{\phi_i}\bigr) \,.
\end{equation}

Finally, using the action variables from \eqref{av} and \eqref{W} we have the angle 
variables (see the Appendix):
\begin{equation}\label{angleti}
\begin{split}
w_{\theta_i} =& \frac{\partial W}{\partial J_{\theta_i}}=
\frac{\partial W_{\theta_i}}{\partial J_{\theta_i}}\\
=& -\frac{J_{\theta_i} + J_{\phi_i}}{2\pi} I_1(a_i,b_i,c_i;\cos \theta_i)
\end{split}
\end{equation}
\begin{equation}\label{anglephii}
\begin{split}
w_{\phi_i} =& \frac{\partial W}{\partial J_{\phi_i}}=
\frac{\partial W_{\theta_i}}{\partial J_{\phi_i}} +
\frac{\partial W_{\phi_i}}{\partial J_{\phi_i}}=
\frac{\partial W_{\theta_i}}{\partial J_{\phi_i}} + \frac1{2\pi}{\phi_i}\\
=&-\frac{J_{\theta_i} + J_{\phi_i}}{2\pi} I_1(a_i,b_i,c_i;\cos\theta_i)\\
&+\frac{-2 J_{\phi_i} + J_\nu}{8\pi} \\
&\times I_2(a_i+b_i +c_i,b_i +2c_i,c_i;\cos\theta_i -1)\\
&+\frac{2 J_{\phi_i} + J_\nu}{8\pi} \\
&\times I_2(a_i-b_i +c_i,b_i -2c_i,c_i;\cos\theta_i +1)\\
&+\frac1{2\pi}\phi_i 
\end{split}
\end{equation}
\begin{equation}\label{anglenu}
\begin{split}
w_{\nu} =& \frac{\partial W}{\partial J_{\nu}}=
\frac{\partial W_{\theta_1}}{\partial J_{\nu}} +
\frac{\partial W_{\theta_2}}{\partial J_{\nu}} +
\frac{\partial W_{\nu}}{\partial J_{\nu}} \\
=&
\frac{\partial W_{\theta_1}}{\partial J_{\nu}} +
\frac{\partial W_{\theta_2}}{\partial J_{\nu}} + \frac1{2\pi}{\nu}\\
=&\sum_{i=1,2}\frac{2 J_{\phi_i} - J_\nu}{16\pi} \\
&\times I_2(a_i+b_i +c_i,b_i +2c_i,c_i;\cos\theta_i -1)\\
&+\sum_{i=1,2}\frac{2 J_{\phi_i} + J_\nu}{16\pi} \\
&\times I_2(a_i-b_i +c_i,b_i -2c_i,c_i;\cos\theta_i +1)\\
&+\frac1{2\pi}\nu \,.
\end{split}
\end{equation}

\section{Conclusions}

Motivated by the great interest of higher order symmetries and their applications
in various field theories, in this paper we describe the construction of conserved
quantities for geodesic motions on toric SE manifold $T^{1,1}$. The integrability 
of geodesics permits us to give an action-angle formulation of the phase space
variables. Such a formulation represents a useful tool for developing of perturbation
theory.

The KAM theorem describes how an integrable system reacts to small perturbations.
The loss of integrability is characterized by the presence of resonant tori which
have rational ratios of frequencies
\begin{equation}
m_i \omega_i = 0 \quad \text{with} \quad m_i\in \mathbb{Q} \,.
\end{equation}
That is the case of the frequencies \eqref{freq} which evinces that the frequencies 
corresponding to $\theta_i$ and $\phi_i$ coordinates are equal.

This result is in accord with the numerical simulations \cite{BZ1} which show that
certain classical string configurations in $AdS \times T^{1,1}$ are chaotic. It is 
considered a string localized on the center of $AdS_5$ that wraps the directions 
$\phi_1$ and $\phi_2$ in $T^{1,1}$. To numerically investigate the perturbation of the 
integrable Hamiltonian by a small non-integrable piece, Ref. \cite{BDG} considered 
Poincar\'{e} sections. In \cite{BZ2} it is presented an analytical approach 
of the loss of the integrability using the techniques of differential Galois theory 
for \emph{normal variational equation}. It is quite remarkable that while the 
point-like string (geodesic) equations are integrable in some backgrounds, the 
corresponding extended classical string motion is not integrable in general 
\cite{BZ2,AKKY}. A similar situation encountered in \cite{CL} in the study of 
(non)-integrability of geodesics in $D$-brane background.

Since the Reeb vector field $K_{\eta}$ \eqref{Rv} is nowhere vanishing, its orbits 
define a foliation of the SE space $M$. If all the orbits close, $K_{\eta}$
generates a circle action on $M$. If moreover the action is free the SE manifold is 
said to be \emph{regular} and is the total space of a $U(1)$ principle bundle over a 
K\"{a}hler-Einstein space. That is the case of the $T^{1,1}$ space with the Reeb vector
field \eqref{TRv}. More generally the $U(1)$ action on $M$ is only locally free 
and such structures are referred to as \emph{quasi-regular}. The geometries $Y^{p,q}$
with $4p^2 - 3 q^2$ a square are examples of such manifolds. If the orbits of the 
Reeb vector field do not close the Kv generates an action $\mathbb{R}$ 
on $M$, with the orbits densely filling the orbits of a torus and the SE
space is said to be \emph{irregular}. The manifolds $Y^{p,q}$ with $4p^2 - 3 q^2$ not
a square represent examples of such geometries \cite{MS}.

From these considerations it is obvious  that there are significant differences between 
$T^{1,1}$ and $Y^{p,q}$ SE spaces. The construction of the action-angle 
variables for integrable geodesic motions in quasi-regular  and irregular spaces is more 
intricate and deserve a special study \cite{MVprep}.

Concluding, KY tensors provide a powerful tool for studying symmetries of black holes 
in higher dimensions and stringy geometries. It would be interesting to extend the 
action-angle formulation to quasi-regular and irregular $5$-dimensional SE metrics
and their higher dimensional generalizations relevant for the predictions of the 
AdS/CFT correspondence.

\section*{Acknowledgements}
This work has been partly supported by the project {\it CNCS-UEFISCDI 
PN--II--ID--PCE--2011--3--0137} and partly by the project {\it NUCLEU 16 42 01 01/2016}.

\section*{Appendix}

{\bf Calculation of the integrals from eqs. \eqref{angleti}--\eqref{anglenu}}

Putting $\cos\theta_i =t_i$ in \eqref{Wti} the characteristic function $W_{\theta_i}$ 
is given by the integral:
\begin{equation}\label{intW}
W_{\theta_i} = \frac1{2\pi}\int \frac{dt_i}{t^2_i-1}\sqrt{a_i + b_i t_i + c_it^2_i}
\end{equation}
with:
\begin{equation}\label{abc}
\begin{split}
a_i &=  J^2_{\theta_i} + 2J_{\theta_i}J_{\phi_i} - \frac14 J^2_{\nu}\\
b_i &= J_{\phi_i}J_{\nu}\\
c_i &= - \bigl( J_{\theta_i} +  J_{\phi_i}\bigr)^2\,.
\end{split}
\end{equation}

For the evaluation of angle variables we need the partial derivatives
of $W_{\theta_i}$ \eqref{intW} with respect to action 
variables $J_{\theta_i}, J_{\phi_i}$ and $J_{\nu}$.

First we evaluate 
\begin{equation}
\frac{\partial W_{\theta_i}}{\partial J_{\theta_i}}= 
-\frac{J_{\theta_i} + J_{\phi_i}}{2\pi} \int \frac{dt_i}{\sqrt{a_i + 
b_it_i + c_it^2_i}}\,.
\end{equation}
Taking into account that in \eqref{abc} $c_i < 0$ the last integral is \cite{GR}
\begin{equation}
I_1(a_i,b_i,c_i;t_i) = \frac{-1}{\sqrt{-c_i}}\arcsin 
\Biggl(\frac{2 c_i t_i + b_i}{\sqrt{-\Delta_i}}\Biggr)
\end{equation}
where $\Delta_i = 4a_ic_i - b^2_i <0$ taking from granted \eqref{constr}.

For the partial derivative of $W_{\theta_i}$ \eqref{intW} with respect to action 
variable $J_{\phi_i}$ we have
\begin{equation}
\begin{split}
\frac{\partial W_{\theta_i}}{\partial J_{\phi_i}}=&
-\frac{J_{\theta_i} + J_{\phi_i}}{2\pi} \int \frac{dt_i}{\sqrt{a_i + b_it_i + c_it^2_i}}\\
&-\frac{J_{\phi_i}}{2 \pi}\int \frac{dt_i}{t^2_i -1}\frac1{\sqrt{a_i + b_it_i + c_it^2_i}}\\
&+\frac{J_\nu}{4 \pi}\int dt_i \frac{t_i}{t^2_i-1}\frac1{\sqrt{a_i + b_it_i + c_it^2_i}}\\
=&-\frac{J_{\theta_i} + J_{\phi_i}}{2\pi} I_1(a_i,b_i,c_i;t_i)\\
&+\frac{-2 J_{\phi_i} + J_\nu}{8\pi} \\
&\times I_2(a_i+b_i +c_i,b_i +2c_i,c_i;t_i -1)\\
&+\frac{2 J_{\phi_i} + J_\nu}{8\pi} \\
&\times I_2(a_i-b_i +c_i,b_i -2c_i,c_i;t_i +1) 
\end{split}
\end{equation}
where \cite{GR}
\begin{equation}
\begin{split}
I_2(a,b,c;t) =&\int \frac{dt}{t\sqrt{a + bt + ct^2}}\\
=&\frac{1}{\sqrt{-a}}\arctan \Biggl(\frac{2a +bt}{2\sqrt{-a}
\sqrt{a +bt +ct^2}}\Biggr)
\end{split}
\end{equation}
for $a<0$.

Finally for the partial derivative of $W_{\theta_i}$ \eqref{intW} with respect to action 
variable $J_{\nu}$ we have
\begin{equation}
\begin{split}
\frac{\partial W_{\theta_i}}{\partial J_\nu}=&
\frac{J_{\phi_i}}{4 \pi}\int dt_i \frac{t_i}{t^2_i -1}\frac1{\sqrt{a_i + b_it_i + c_it^2_i}}\\
&-\frac{J_\nu}{8 \pi}\int \frac{dt_i}{t^2_i-1}\frac1{\sqrt{a_i + b_it_i + c_it^2_i}}\\
=&\frac{2 J_{\phi_i} - J_\nu}{16\pi} I_2(a_i+b_i +c_i,b_i +2c_i,c_i;t_i -1)\\
&+\frac{2 J_{\phi_i} + J_\nu}{16\pi} \\
&\times I_2(a_i-b_i +c_i,b_i -2c_i,c_i;t_i +1)\,.
\end{split}
\end{equation}


\begin{thebibliography}{99}

\bibitem{JS}
J. Sparks, 
Surv. Diff. Geom. \textbf{16}, 265 (2011)

\bibitem{GMSW}
J. P. Gauntlett, D. Martelli, J. Sparks, D. Waldram,
Adv. Theor. Math. Phys.  \textbf{8}, 987 (2004)

\bibitem{CLPP}
M. Cveti\v c, H. L\"{u}, D. N. Page, C. N. Pope,
Phys. Rev. Lett. \textbf{95}, 071101 (2005)

\bibitem{NB}
N. Beisert, et al,
Lett. Math. Phys. \textbf{99}, 3 (2012)

\bibitem{BPR}
I. Bena, J. Polchinski, R. Roiban,
Phys. Rev D \textbf{69}, 046002 (2004)

\bibitem{BZ2}
P. Basu, L. A. Pando Zayas,
Phys. Rev. D \textbf{84}, 046006 (2011)

\bibitem{KW} 
I. R. Klebanov,  E. Witten,
Nucl. Phys. B \textbf{536}, 199 (1988)

\bibitem{VIA}
V.I. Arnold,
\textit{Mathematical Methods of Classical Mechanics}, 2nd edn.
(Springer-Verlag, New York, 1989)

\bibitem{US}  
U. Semmelmann,  
Math. Z. \textbf{245}, 503 (2003) 

\bibitem{B-G-2008}
C. Boyer, K. Galicki,
\textit{Sasakian geometry},
(Oxford Mathematical Monographs, Oxford University Press, Oxford, 2008)

\bibitem{CO} 
P. Candelas, X. C. de la Ossa, 
Nucl. Phys. B \textbf{B342}, 246 (1990)

\bibitem{MS}  
D. Martelli, J. Sparks, 
Commun. Math. Phys. \textbf{262}, 51 (2006)

\bibitem{BV}
E. M. Babalic, M. Visinescu,
Mod. Phys. Lett. A \textbf{30}, 1550180 (2015)

\bibitem{S-V-V1}
V. Slesar, M. Visinescu, G. E. V\^ilcu, 
EPL \textbf{110}, 31001 (2015)

\bibitem{S-V-V2}
V. Slesar, M. Visinescu, G. E. V\^ilcu,
Annals of Physics \textbf{361}, 548 (2015)

\bibitem{GPS}
H. Goldstein, C. Poole, J. Safko,
\textit{Classical Mechanics}, 3rd edition
(Addison-Wesley, San Francisco, 2002)

\bibitem{BZ1}
P. Basu, L. A. Pando Zayas,
Phys. Lett. B \textbf{700}, 243 (2011)

\bibitem{BDG}
P. Basu, D. Das, A. Ghosh,
Phys. Lett. B \textbf{699}, 388 (2011)

\bibitem{AKKY}
Y. Asano, K. Kawai, H. Kyono, K. Yoshida,
JHEP \textbf{08}, 060 (2015)

\bibitem{CL}
Y. Chervonyi, O. Lunin,
JHEP \textbf{02}, 061 (2014)

\bibitem{MVprep}
M. Visinescu, work in progress

\bibitem{GR}
I.S. Gradshteyn, I.M. Ryzhik, 
\textit{Table of Integrals, Series, and Products}, ed. by A. Jeffrey and 
D. Zwillinger, 7th edn. (Academic Press, New York, 2007)

\end{thebibliography}
\end{document}